\documentclass[pra,showpacs,twocolumn,aps]{revtex4-1}

\usepackage[utf8]{inputenc}
\usepackage{latexsym}
\usepackage{amssymb}
\usepackage{amsfonts}
\usepackage{epsfig}
\usepackage{psfrag}
\usepackage{graphicx}

\usepackage{epsfig,color}
\usepackage{psfrag}
\usepackage[utf8]{inputenc} 
\definecolor{darkgreen}{rgb}{0,0.4,0}

\newcommand{\f}[1]{\mbox{\boldmath$#1$}}

\newcommand{\bea}{\begin{eqnarray}}
\newcommand{\ea}{\end{eqnarray}}
\newcommand{\eea}{\end{eqnarray}}

\newcommand{\vc}[1]{\mathbf{#1}}

\usepackage{amsmath}
\usepackage{longtable}
\usepackage{color}

\begin{document}

\title{Screened superexchange mechanism for superconductivity applied to cuprates}
\author{Patrick Navez$^{1,2}$}
\affiliation{$^1$ Laboratoire Charles Coulomb UMR 5221 CNRS-Universit\'e de Montpellier, F-34095 Montpellier, France \\
$^2$ Department of Physics, Loughborough University,
Loughborough LE11 3TU, United Kingdom
}

\date{\today}
\begin{abstract}
In 1965, Kohn and Luttinger published a note revealing that dynamical screening of the repulsive Coulomb interaction leads under certain conditions to an effective attraction necessary for the formation of Cooper pairs. We propose such a formalism adapted to the cuprates where the screening arises from the superexchange dynamics of virtual holes in the oxygen orbitals of the $Cu O_2$ plane. Using an adequate Schrieffer-Wolff transformation, the basic Hartree-Fock-Bogoliubov (HFB) method and the {\it ab initio} data on orbitals (energy, hopping, interaction), we derive some predictions for the temperature-doping phase diagram (pseudo-gap, strange metal, antiferromagnetism, superconducting and normal states)  and for the doping dependant band energy spectrum  in semi-quantitative agreement with observations.
\end{abstract}
\maketitle


\section{Introduction}

High-temperature superconductivity in cuprates has been the subject of considerable theoretical investigation over the past four decades, yet no widely accepted approach has emerged \cite{RevModPhys.95.021001,Leggett}. The basic  Bardeen–Cooper–Schrieffer (BCS) theory based on electrons (holes) pairing caused by  the phonon exchange mechanism fails to explain such high transition temperature. Experiments on isotope effects suggest rather that the phonons do not play major role for  superconductivity in such ceramic materials. Instead, different alternative many body approaches have been studied with the idea that pairing originates from strong correlations between electrons (or holes).
Most of them start from  the  one-band lattice Hubbard Hamiltonian with on-site repulsive interaction  derived from the  more fundamental {\it ab initio } description  \cite{PhysRevB.44.7504,PhysRevB.71.134527,PhysRevB.53.8751,PhysRevB.45.7959}.
These strategies have resulted in many new techniques (DMFT, DFT,  RVB, ...  ) \cite{PhysRevResearch.2.023172,Weber_2010,Baeriswyl_2009,doi:10.1126/science.235.4793.1196,PhysRevLett.60.2430,PhysRevB.81.064515} with some mitigated success, since the validity of the approximations behind this derivation is usually taken for granted and hardly reconsidered, despite the numerous assumptions made. A more systematic derivation based on a controlled  expansion of a small parameter would enforce the Hubbard Hamiltonian or alternatively would lead to a more elaborated and richer Hamiltonian.

In this paper, we shall not review the extensive works developed over the years, but instead focus on the earlier studies from the 1960s with the aim to explain this unconventional superconductivity. The guiding principle is to return to methods that initially proved successful, before turning to newer developments, with the emphasis that many early many-body approaches may have been abandoned without thorough examination of their full potential.

The proposed methodology is inspired by the proposal of Kohn and Luttinger which states that pairing could  originate from the dynamical screening of the bare repulsive Coulomb interaction between two electrons leading to an effective attraction at a certain distance \cite{KL,LIU199781,Belyavsky}. In a simple picture, the other surrounding electrons obeying the Pauli exclusion principle contribute to alter the profile of the two-body interaction. In the case of cuprates, we shall show that the inclusion of the screening to the superexchange mechanism in the $Cu O_2$ plane leads to a Hamiltonian with an effective attraction between holes \cite{Anderson,ZR,ZR2}.

For this purpose, the main established techniques deployed here are the following:
\begin{enumerate}
 \item We start from a lattice Hamiltonian describing the main orbitals of the electron in the two-dimensional $Cu O_2$ plane. In the absence of interaction, it consists in a simple tight-binding model with $\sigma$ bonds between the orbitals $3 d_{x^2-y^2}$ for the copper ($d$)  and $p_{x,y}$ for the oxygen ($p$) resulting in three energy bands for holes. The parameters of the lattice model can be extracted from an {\it ab initio} approach \cite{Annett,PhysRevB.39.9028,ANDERSEN19951573}. More modern works exist based on an improved  density functional method  for all atoms involved in the cuprates including those responsible for doping (see for instance \cite{doi:10.1073/pnas.1910411116}). But
 we shall restrict ourselves only to the orbitals responsible for the holes transport.
The other binding orbitals $3 d_{3z^2-r^2}$ mixing with the  orbitals of the two apical oxygens and with the planar orbitals in \cite{Annett} or the involvement of the $s$ orbitals of $Cu$ in \cite{MISHONOV2025417509}
are not relevant for our concerns.
 
 \item We use the Schrieffer-Wolff (SW) unitary transformation in order to rewrite the lattice Hamiltonian in a Fock basis where correlation becomes negligible. Unlike in the BCS theory where the electron-phonon coupling is used as the expansion parameter, we choose for the lattice the hopping parameters $t_{pd}$ between the $Cu$ and $O$ orbitals. This procedure is an alternative but more systematic path to derive the one-band Hubbard model
 \cite{PhysRevB.44.7504,PhysRevB.71.134527,PhysRevB.53.8751,PhysRevB.45.7959}. 

 \item 
 We use the Hartree-Fock-Bogoliubov (HFB) mean field {\it ansatz} as in the BCS theory to  analyze the various phases in cuprates that break the symmetry. According to the Mermin-Wagner theorem, these phases should only appear in three dimensions. To justify the ansatz used in two dimensions, we assume the existence of an effective weak hopping between the $CuO_2$ planes
 but which will be neglected in subsequent calculations
 \cite{Laughlin}.
 \end{enumerate}

Using this theoretical scheme, the resulting mean field model shows that the repulsive Coulomb interaction energy $U_{pd}$ between the holes of $p$ and $d$ orbitals, rarely considered in the literature \cite{Hansmann2014,Zegrodnik_2021}, is crucial for the  effective attraction in the  $d$-wave channel \cite{Leggett,Annett1999}. The presence of additional holes occupying the $d$ orbitals leads to screening and changes the sign of the pairing interaction  under some conditions.

We are also able to describe the pseudogap (PG) phase \cite{doi:10.1073/pnas.1209471109,osti_1488523,TomTimusk_1999} that results from  correlation between adjacent holes with different spins. We define an order parameter breaking the spin symmetry between neighbouring sites, different  from the ones proposed in \cite{PhysRevResearch.2.023172}, in which the hole has  frustrated dynamics, in the sense that it tilts its spin when hopping to an adjacent site. Our derived model also suggests that the strange metal state distinguishes from the Fermi liquid  both by a narrow energy  band and by a singular density of state. Finally, we determine an antiferromagnetic spin coupling $J$ \cite{ZR,Leggett} in agreement with experiments. As a result, we obtain the universal phase diagram of temperature versus doping in semi-quantitative agreement with experiments on cuprates.

In this work, we focus only on the leading order in the hopping parameters, as a proof of principle to highlight the main features of the ideas presented. However, we emphasize that the method developed is quite general and can be extended to higher orders for improved accuracy, as well as to other unconventional superconductors, including different types of orbitals and potentials.

The paper is organized as follows. Starting from the three-band model in Sec. 2, Sec 3 is devoted to the screened superexchange  mechanism and its implication on superconductivity. Sec. 4 discusses how this formalism describes the other phases (pseudogap, strange metal, antiferromagnetism), before ending with the conclusion in Sec. 5. Appendices contain a detailed account of the full formalism using the SW and HFB techniques and some details of the calculations.

\section{Hamiltonian for the three-band model}

We start with a many-body hole description (not electrons) in the $Cu O_2$ plane of the $3 d_{x^2-y^2}$ state ($d$) interacting
with the $2 p_{x,y}$  states ($p$). In absence of doping the $d$ state is half filled, while the
$p$ states are empty (or doubly occupied with electrons) \cite{PhysRevB.44.7504,PhysRevB.71.134527,PhysRevB.53.8751,PhysRevB.45.7959}. 
We label an elementary cell  by $\vc{l}=l_x\vc{1}_x+ l_y\vc{1}_y$ with $l_i = 0,\dots, L-1$ where $N=L^2$ is the site number and $i=x,y$ ($\overline{i}=y,x$).
As usual, we consider a system with periodic boundary conditions. For each cell, we define the fermion creation and annihilation operators for the holes $\hat d^\dagger_{\vc{l},\sigma}$,
$\hat d_{\vc{l},\sigma}$ in the $d$ state and $p^\dagger_{\vc{l}+\vc{1}_i/2,\sigma}$,
$\hat p_{\vc{l}+\vc{1}_i/2,\sigma}$ in the $p$ state with half spin 
$\sigma=\pm 1$ or $\uparrow, \downarrow$.
We use the number operator notation $\hat n_{p, \vc{l}+\vc{1}_i/2,\sigma}=\hat p^\dagger_{\vc{l}+\vc{1}_i/2,\sigma} \hat p_{\vc{l}+\vc{1}_i/2,\sigma}$ and $\hat n_{d, \vc{l},\sigma}=\hat d^\dagger_{\vc{l},\sigma} \hat d_{\vc{l},\sigma}$. In a tight binding description,
the holes travel between the $p$ and $d$ states with a hopping energy  $t_{pd}$ ($1-2  eV$) resulting from a $\sigma$ bonding. For each orbital $p$ and $d$,
we define the on-site repulsion energy
$U_p$ ($  2-5 eV$), $U_d$  ($ \sim 8eV$), and the nearest neighbour one $U_{pd}$  ($0.8-2 eV$).  The  unoccupied orbitals  $p$ have a energy   higher than $d$ state by the amount $\epsilon_p$ ($2 - 4.8 eV$) and are bonded each other with the hopping energy $t_{pp}$ ($\sim 0.3 eV$). The typical values between parenthesis of these energy parameters are indicative and were obtained from {\it ab initio} modelling
\cite{PhysRevB.44.7504,PhysRevB.71.134527,PhysRevB.53.8751,PhysRevB.45.7959}. 

The  Hamiltonian operator has the expression:
\begin{eqnarray}\label{H1}
\hat H_T &=& \hat H_0 + \hat V  \, ,
\end{eqnarray}
where
\begin{widetext}
\begin{eqnarray}\label{H2}
\hat H_0&=&
\sum_{\vc{l},\sigma,i=x,y}
 \hat n_{p,\vc{l} + \vc{1}_i/2,\sigma}\left[\epsilon_p + U_{pd} \sum_{\sigma'} (\hat n_{d,\vc{l},\sigma'}+ \hat n_{d,\vc{l}+ \vc{1}_i,\sigma'})\right]
+
\sum_{\vc{l}} U_d \hat n_{d,\vc{l},\uparrow}\hat n_{d,\vc{l},\downarrow}+\sum_{\vc{l},i=x,y} U_p
\hat n_{p,\vc{l} + \vc{1}_i/2,\uparrow}\hat n_{p,\vc{l} + \vc{1}_i/2,\downarrow}
\nonumber\\
&+&\sum_{\vc{l},\sigma,i=x,y} t_{pp} \left(\hat p^\dagger_{\vc{l} +\vc{1}_{i}/2,\sigma} + \hat p^\dagger_{\vc{l} +\vc{1}_{\overline i}+\vc{1}_i/2,\sigma}\right)\left(\hat p_{\vc{l}+\vc{1}_{\overline i}/2,\sigma}+\hat p_{\vc{l}+ \vc{1}_i+\vc{1}_{\overline i}/2,\sigma}\right)
\\ \label{H3}
\hat V&=&
\sum_{\vc{l},\sigma,i=x,y} t_{pd} \left[\hat p^\dagger_{\vc{l} +\vc{1}_i/2,\sigma} (\hat d_{\vc{l},\sigma}+\hat d_{\vc{l}+ \vc{1}_i,\sigma}) +
(\hat d^\dagger_{\vc{l},\sigma}+\hat d^\dagger_{\vc{l}+ \vc{1}_i,\sigma})
\hat p_{\vc{l} + \vc{1}_i/2,\sigma} \right] \,.
\end{eqnarray}
\end{widetext}
The perturbation term $\hat V$ is controlled by the expansion parameters $t_{pd}$ and is used in the SW transformation in 
appendix $\ref{A}$. 


\section{Proposed superconductivity mechanism}

\subsection{Perturbation approach on three sites}

The proposed mechanism for superconductivity can be explained in terms of the simpler triad cluster of holes occupying two $d$ sites $\vc{l}$ and $\vc{l}+\vc{1}_i$ and one $p$ site in between with two bonds only. We calculate the ground state energy from a perturbation approach for the case of no hole $u_0$, one hole $u_1$ and $u_{1'}$, or two paired holes $u_2$ at different sites. As seen in table \ref{t:1}, up to the  second order in $t_{pd}$, there is a virtual hopping of one isolated hole to the $p$ orbital lowering the total energy as much as the energy transition is smaller and causing some delocalization of the holes from the $d$ orbital into the $p$ orbital. On the contrary for virtual transitions involving double occupancy, the shift in energy is higher but globally less than the case of an isolated hole for sufficiently high on-site repulsive interaction $U_d$.

\begin{figure}
 \centering
 \includegraphics[width=8cm]{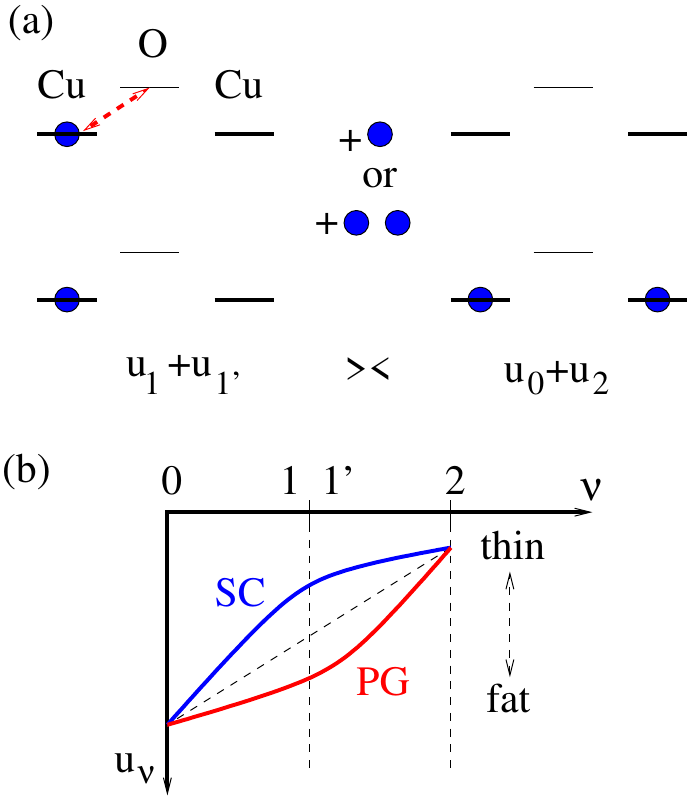}
 \caption{Schematic explanation of pairing:
 (a) Four independant triads consisting each of two $d$ sites interacting via a $p$ site. Comparison between the energy of two single holes  $u_1 +u_{1'}$ and the energy of no hole and pairing of energy $u_0+u_2$. (b)
 Schematic energy graph of $u_\nu$ for various occupation $\nu$.
 The repulsive potential $U_{pd}$ is crucial for pairing. Its screening role reduces the  delocalisation of any hole on the $p$ orbital and increases the energy. Such a ``thin'' hole contrasts with a ``fat'' hole with a more delocalized state on the $p$ orbital.
 SC pairing becomes favorable with one additional hole where $u_\nu$ takes a convex form caused by the perturbative term $-t_{pd}^2/(\epsilon_p + \nu U_{pd})$.
 Similarly, $u_\nu$ is also convex for two additional holes but concave in absence of any hole leading to the PG phase.
 }
 \label{fig:0}
\end{figure}

Pairing occurs if the presence of two holes is more favorable energetically than two isolated holes relative to the background energy as shown in Fig.\ref{fig:0}. The resulting attraction energy should be negative: $u_2 +u_0-u_1-u_{1'} < 0$. As seen in table \ref{t:1}, this energy is positive unless additional holes generate screening. In the latter case, the transition energy is reduced by the presence of repulsion energy $U_{pd}$ leading to a global net attraction.
Adding one hole, attraction is possible in the limit of strong
on-site repulsive interaction $U_d$ with a quadratic contribution in  $U_{pd}$. Only in the case of two added holes, attraction is possible with a linear contribution in $U_{pd}$. As a main result of this work, we conclude that  virtual holes to the $p$ orbital occur conditionally on the presence of the  other surrounding holes and this process transforms a repulsive Coulomb force into an effective attraction for some parameter values.

The pairing is also influenced by the hopping energy
$u_{t,\nu}=-t_{pd}^2/\epsilon_\nu$ between the adjacent sites with a transition energy $\epsilon_\nu$ from  a $d$ orbital to the $p$ orbital as seen from the table \ref{t:2}. The presence of other holes in the $d$ sites modifies this transition energy by screening.
The total attraction energy is $U_{t2}=(-1)(u_{t,2} -u_{t,1} -u_{t,1'} +u_{t,0})$ with a statistical factor $(-1)$ that accounts for the hole spin exchange. Any hopping must indeed satisfy the exclusion principle in a pairing with opposite spin.


\begin{widetext}

\begin{table}[h!]
\centering
\begin{tabular}{|c|c| c| c |c| c|c|}
\hline
Occupancy &
\multicolumn{2}{|c|}{Cu}& O& \multicolumn{2}{|c|}{Cu}& Potential Energy/pair   $u_0$, $u_1$, $u_{1'}$, $u_2$\\
\hline
$(1-n)^2$
&0 &0 & & 0 &0& $0$ \\
&0 &1 & & 0 &0& $-t_{pd}^2/\epsilon_p$ \\
&0 &0 & & 1 &0& $-t_{pd}^2/\epsilon_p$ \\
&0 &1 & & 1 &0& $-2 t_{pd}^2/(\epsilon_p +U_{pd})$ \\
\hline
\multicolumn{7}{|c|}{$\displaystyle u_{a,0}=t_{pd}^2\left(\frac{2 }{\epsilon_p}-\frac{2}{\epsilon_p +U_{pd}} \right)$}\\
\hline
$n(1-n)$
&1 &0 & & 0 &0& $-t_{pd}^2/\epsilon_p$ \\
&1 &1 & & 0 &0& $U_d - 2t_{pd}^2/(\epsilon_p-U_d + U_{pd})$ \\
&1 &0 & & 1 &0& $-2t_{pd}^2/(\epsilon_p+U_{pd})$ \\
&1 &1 & & 1 &0& $U_d - t_{pd}^2/(\epsilon_p +2 U_{pd}) - 2 t_{pd}^2/(\epsilon_p +2 U_{pd} -U_d)$ \\
\hline
\multicolumn{7}{|c|}{
$\displaystyle u_{a,1}=t_{pd}^2\left(\frac{2 }{\epsilon_p + U_{pd}-U_d}-\frac{2}{\epsilon_p + 2 U_{pd}-U_d}-\frac{1}{\epsilon_p}+ \frac{2}{\epsilon_p+ U_{pd}} -\frac{1}{\epsilon_p+ 2U_{pd}}\right)$ }\\
\hline
$n^2$
&1 &0 & & 0 &1& $-2t_{pd}^2 /(\epsilon_p+U_{pd})$ \\
&1 &1 & & 0 &1& $U_d - t_{pd}^2/(\epsilon_p +2 U_{pd}) - 2 t_{pd}^2/(\epsilon_p +2 U_{pd} -U_d)$ \\
&1 &0 & & 1 &1& $U_d -  t_{pd}^2/(\epsilon_p +2 U_{pd}) -  2 t_{pd}^2/(\epsilon_p +2 U_{pd} -U_d)$ \\
&1 &1 & & 1 &1& $2 U_d - 4 t_{pd}^2/(\epsilon_p +3 U_{pd}- U_d) $ \\
\hline
\multicolumn{7}{|c|}{
$\displaystyle u_{a,2}=t_{pd}^2\left(\frac{4}{\epsilon_p + 2 U_{pd}-U_d}- \frac{4}{\epsilon_p + 3 U_{pd}-U_d}  + \frac{2}{\epsilon_p+ 2U_{pd}} -\frac{2}{\epsilon_p+ U_{pd}}\right)$ }\\
\hline
\end{tabular}
\caption{Determination of the energy for vacuum, single or double occupation according to the screening of $0$,$1$ or $2$ additional holes.
Calculation of the corresponding attraction energy $u_{a,0}$, $u_{a,1}$ and $u_{a,2}$.  }
\label{t:1}
\end{table}

\begin{table}[h!]
 \centering
\begin{tabular}{|c|c| c| c |c| }
\hline
Hopping &
Cu& O& Cu& Hopping Energy/pair   $u_{t,0}$, $u_{t,1}$, $u_{t,1'}$, $u_{t,2}$\\
\hline
$n_\epsilon$
&0 & & 0 & $-t_{pd}^2/\epsilon_p$ \\
&1 & & 0 & $-t_{pd}^2/(\epsilon_p-U_d + U_{pd})$ \\
&0 & & 1 & $-t_{pd}^2/(\epsilon_p+U_{pd})$ \\
&1 & & 1 & $-t_{pd}^2/(\epsilon_p +2 U_{pd} -U_d)$ \\
\hline
\multicolumn{5}{|c|}{
$\displaystyle
U_{t2}=
t_{pd}^2\left(\frac{1}{\epsilon_p}-\frac{1}{U_{pd} +\epsilon_p}
+\frac{1}{U_d -U_{pd} -\epsilon_p}-\frac{1}{U_d -2 U_{pd} -\epsilon_p}\right)
$
}\\
\hline
\end{tabular}
\caption{Determination of the hopping energy for vacuum, single or double occupation. Calculation of the attraction energy $U_{t2}$.}
\label{t:2}
\end{table}
\end{widetext}

\subsection{Generalization to the mean field approach}

The simple reasoning obtained so far from three sites 
can be generalized to the full lattice. Inspired from the BCS approach  where the  SW transformation generates a virtual process resulting in an effective  electron Hamiltonian eliminating
the phonon degree of freedom, the same transformation is used to eliminate the  hole degree of freedom in the $p$ orbitals. It results in  a simpler  Hamiltonian  (\ref{H}) with an effective  hopping term between adjacent $d$ orbitals developed in appendix \ref{A}.
We define the average density and the hopping average per site and spin as:
\begin{eqnarray}
n&=& \sum_{\vc{l},\sigma} \frac{\langle \hat n_{d,\vc{l},\sigma}\rangle}{2N} \, , \quad \quad
n_{\epsilon}=\sum_{\vc{l},\sigma,\pm,i=x,y} \frac{\langle \hat d^\dagger_{\vc{l},\sigma}\hat d_{\vc{l}\pm \vc{1}_{i},\sigma}\rangle}{8N} \,.
\end{eqnarray}
Here, we use the notation $n$ for doping instead of $p=2n-1$.
The absence of doping or half-filling corresponds to the density per spin $n=1/2$ where  $n_\epsilon = 2/\pi^2$ has  its highest value in the Hartree-Fock (HF) approximation  (see Eq.(\ref{n}) and Eq.(\ref{ne})  for explicit expressions).
In our leading order description, we neglect the effective hopping between the next to nearest neighbouring $d$ sites resulting from multiple virtual hopping between the $p$ orbitals.

In appendix \ref{B}, the minimization using the  HFB {\it ansatz} allows to identify an attraction potential $U$ with a pairing corresponding to the order parameter of d-wave symmetry:
\begin{eqnarray}
m&=&\sum_{\vc{l},\pm} \frac{\langle \hat d_{\vc{l},\downarrow}(\hat d_{\vc{l}\pm \vc{1}_{x},\uparrow}-\hat d_{\vc{l}\pm \vc{1}_{y},\uparrow})\rangle}{4N} \,.
\end{eqnarray}
Alternatively, this potential can be deduced more simply by assuming  that the population of the assisting holes is on average the density of particle $n$. 
From the table \ref{t:1}, we deduce by weighting the occupancy that:
\begin{eqnarray}
U_{a}&=&(1-n)^2 u_{a,0} +  2(1-n)n\, u_{a,1} + n^2 u_{a,2} \,.
\end{eqnarray}
From the table \ref{t:2}, we deduce the exchange hopping energy as the second contribution:
\begin{eqnarray}
U_{b}&=& 4 U_{t2}n_\epsilon \,.
\end{eqnarray}
The factor 4 corresponds to all possible directions of hopping along the $x$ and $y$ axis.
Therefore, the total potential is $U=U_a +U_b$.
In the more formal calculation of appendix \ref{B},
another  contribution $n_\epsilon^2$ arises to the quadratic term $n^2 \rightarrow n^2 +n_\epsilon^2$. For any $U_{pd}$ and weak pairing, we obtain from Eq.(\ref{D}) the attractive potential criterion:
\begin{eqnarray}\label{crit}
U
=4[U_{d1}-2U_{d2}n+U_{d3}(n^2+ n_\epsilon^2)+ U_{t2}n_\epsilon] <0 \,,
\end{eqnarray}
where the expressions $U_{di}$ are found in Eqs.(\ref{U}).
From Eq.(\ref{crit}), we note again that $U= 0$ for $U_{pd}=0$ and $n=0$ proving that both repulsive potential between holes of orbitals $p$ and $d$  and  high doping  are the two major ingredients for superconductivity.
Two limit cases deserve a peculiar attention:

\begin{itemize}
 \item
For weak  off-site repulsion $U_{pd}$ and strong on-site repulsion $U_d$ and positive $n_{\epsilon}$, we obtain simply $1- 2n < n_\epsilon$ which is achieved for $1/2+1/\pi^2\simeq 0.6 <n<1$. 
In addition, the resulting attraction is too weak for an observable
significant pairing.

\item
In the limit of weak charge transfer energy $\epsilon_p \rightarrow 0$, then the hopping average has to be negative $n_\epsilon < 0$ in order to guarantee the right sign for the potential. We obtain:
\begin{eqnarray}
n_\epsilon+2n^2-3n+1
+2n_\epsilon^2 < 0
\end{eqnarray}
This regime is a universal free parameter case and is attractive for
$0.4 <n< 1$. It corresponds to a model Hamiltonian that restrains
hopping unless the input and output sites are both empty.
\end{itemize}

The second case appears to be more realistic to describe cuprates as the possibilities of attractive potential are larger.

\subsection{Hopping term}

As seen previously, the sign of the hopping average $n_\epsilon$ determines the conditions for which superconductivity occurs.
In a second  order description and choosing $t_{pp}=0$, only the nearest neighbor hopping contributes to the kinetic of hole with the dispersion energy
$\epsilon_{\vc{k}}=-2t(\cos k_x + \cos k_y )$ in the lattice space of wavevector $\vc{k}=(k_x,k_y)$.
For the free lattice gas, the energy per spin and hole has the form $-4 t_0 n_\epsilon$ with a constant prefactor $t=t_0=t_{pd}^2/\epsilon_p$ \cite{ZR}. Instead, in the HF approximation, the
structure of the superexchange leads to an effective hopping Eq.(\ref{t}) that
depends on $n$ and $n_\epsilon$ under the polynomial form:
\begin{eqnarray}
t&=&t_0-U_{t1}n +U_{t2}
(n^2 -3n_\epsilon^2)
\nonumber \\ 
&+&[2U_{d1}-4U_{d2}n+ 2U_{d3}(n^2-n_\epsilon^2)]n_\epsilon \,.
\end{eqnarray}
The choice of a negative sign for
$n_\epsilon$ for an effective attraction imposes also a negative $t$.
For comparison at half-filling for $U_{pd}=0$ and large $U_d$, the hopping term is $t_{1/2} \simeq t_0/2$ which differs by a factor $1/2$  from the one used in the literature \cite{ZR}.

\begin{figure}
 \centering
 \includegraphics[width=9cm]{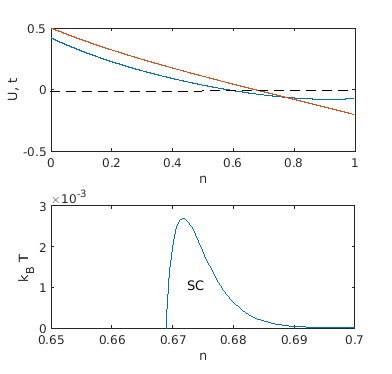}
\caption{First graph of the hopping factor $t$ in red and potential $U$ in blue vs. the doping $n$. The dash line on the x axis is a guide for the eye.  Second graph of the critical superconducting (SC) temperature vs. the doping $n$ for the parameter values $\epsilon_p=2$, $U_{pd}=0.53$, $U_d=7.925$ and  $t_{pp}=0$ expressed in unit of $t_{pd}$.
}
 \label{fig:1}
\end{figure}

\begin{figure}
 \centering
 \includegraphics[width=9cm]{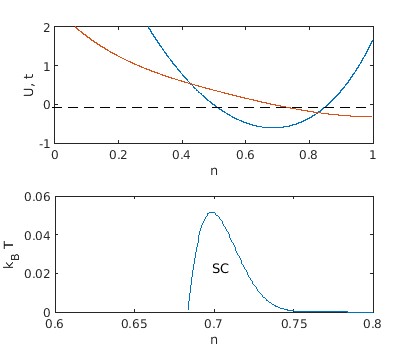}
 \caption{Same as in Fig.\ref{fig:1}  but for $\epsilon_p=0.4$, $U_{pd}=1.55$, $U_d=6.6$ and  $t_{pp}=0$.}
 \label{fig:2}
\end{figure}

\begin{figure}
 \centering
 \includegraphics[width=9cm]{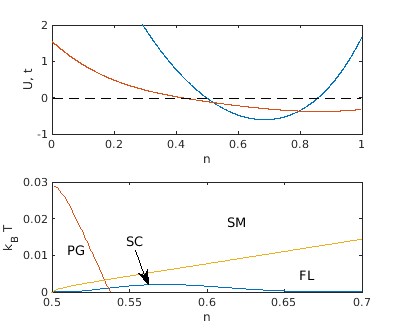}
 \caption{Same as in Fig.\ref{fig:1}  but for $\epsilon_p=0.4$, $U_{pd}=1.55$, $U_d=6.6$ and $t^{eff}_{pp}=0.038$ and with the transition line for the pseudogap (PG) the strange metal (SM) and the Fermi liquid (FL) phases in the second graph.}
 \label{fig:3}
\end{figure}

\begin{figure}
 \centering
 \includegraphics[width=9cm ]{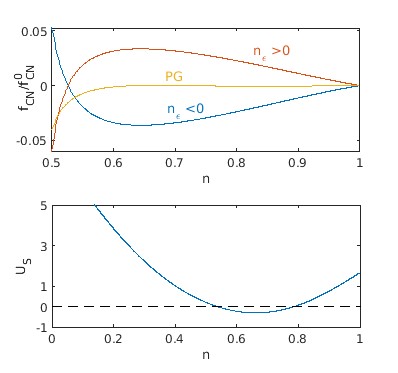}
 \caption{First graph: canonical free energy $f_{CN}$ vs. doping $n$ at zero temperature for the HF state for positive and negative $n_\epsilon$  and for the pseudogap expressed with respect to the free energy reference $f^{0}_{CN}$  without hopping (see  Eq.(\ref{f0})). Second graph:
 pseudogap potential vs. the doping $n$. We use the parameters of Fig.\ref{fig:3}.}
 \label{fig:4}
\end{figure}

In Figs.\ref{fig:1} and \ref{fig:2},
we plot both the variation of $t$ and $U$ with the doping for negative
$n_\epsilon$ in the HF approximation.
The graphs show that a negative $t$ occurs only at high doping $n \sim 0.7$. To achieve it close to half-filling, we need to account for the hopping contribution $t^{eff}_{pp}$ between the $p$ orbitals as in Fig.\ref{fig:3} defined in Eq.(\ref{tppeff}). In the latter case, the canonical (CN) free energy Eq.(\ref{fHF}) is shown in Fig.\ref{fig:4}
and is lower for negative $n_\epsilon$.

\subsection{Critical temperature and energy spectrum}

The critical temperature  $T_{cs}$  beyond which the superconducting phase disappears is determined from the gap equation (see Eq.(\ref{Tc})).
We find the scaling formula :
\begin{eqnarray}\label{Tcc}
k_B T_{cs}\sim 2|t|\exp(- 1/|D_d(\mu) U|)
\end{eqnarray}
where $D_d(\mu)\cong \ln|16t /(e^2 \mu)|/(2\pi^2|t|)$ is the density of state for pairing and $\mu$ is the effective chemical potential defined in (\ref{mu}).
Contrary to the BCS formula that involves phonon exchange, the present expression contains only the parameters of three-band model and does not include any additional attraction. Pairing is genuinely caused by the influence of the repulsive Coulomb interaction $U_{pd}$ in virtual transitions involving the $p$ orbitals.

In Figs.\ref{fig:1},\ref{fig:2},\ref{fig:3}, we represent the variation of critical temperature $T_{cs}$ with the doping. In the limit of flat band for the $p$ orbital ($t_{pp}=0$), high doping is needed to
satisfy both negative hopping and attractive potential, but for significant effective $p$ energy band, pairing becomes possible with doping closer to half-filling.
The vanishing superconductivity for higher hole doping is not caused by the absence of attraction but by the strong reduction of the density of states. The high value of  $T_{cs}$ in Fig.\ref{fig:2}
offers an insight on the role of screening in the search for room temperature superconductors far from half-filling.
In \cite{doi:10.1073/pnas.2207449119}, the charge transfer energy gap decreases when the superconducting order parameter $m$ increases. Although this energy gap is different according to the hole population  involved in screening,
it can be represented by the parameter $\epsilon_p$ for which a lower value increases  the transition temperature  as confirmed from Figs.\ref{fig:1} and \ref{fig:2}.

The energy spectrum obtained from appendix \ref{B} has a BCS-like form:
\begin{eqnarray}
E_{\vc{k}}=\sqrt{(\epsilon_{\vc{k}} - \mu)^2 + |\Delta|^2 (\cos k_x - \cos k_y)^2/4 } \,.
\end{eqnarray}
It displays a gap $\Delta=-Um$ except at the nodal points $k_x=\pm k_y$  in semi-quantitative agreement with the observation from the  ARPES experiment which provides data on the energy band structure  \cite{osti_1488523,TomTimusk_1999,PhysRevLett.104.207002}.

\section{Other phases}

\subsection{Frustrated hole - Pseudogap}

Despite the enormous literature, the nature of pseudogap phase is still yet  unclear
in the scientific community. The screened superexchange formalism
has the merit to propose an explanation to this exotic phase.
When the pairing potential becomes repulsive $U > 0$ (see also Fig.\ref{fig:0}), another $d$-wave like symmetry is spontaneously broken leading to the non trivial order parameter:
\begin{eqnarray}\label{PG}
n_{\overline{\epsilon}}=\sum_{\vc{l},\pm} \frac{\langle \hat d^\dagger_{\vc{l},\downarrow}(\hat d_{\vc{l}\pm \vc{1}_{x},\uparrow}-\hat d_{\vc{l}\pm \vc{1}_{y},\uparrow})\rangle}{4N} \, .
\end{eqnarray}
It has the effect that a hole travelling to the lattice spontaneously tilts its spin at each hopping. 
Therefore the spin of the hole is not a good quantum number anymore and the dispersion relation splits in two branches:
\begin{eqnarray}
E^{PG}_{\pm, \vc{k}}= \epsilon_{\vc{k}} \pm |\Delta_{PG}|\frac{\cos k_x - \cos k_y}{2} \,,
\end{eqnarray}
with a energy gap measured by $\Delta_{PG}=U n_{\overline{\epsilon}}$.

There are two Fermi surfaces obtained by setting $E^{PG}_{\pm, \vc{k}}=\mu$. The surfaces cross at the nodal points  $\vc{k}=(\pm \pi/2,\pm \pi/2)$ and depart maximally at the antinodal region. The Fermi arc reported in \cite{osti_1488523,TomTimusk_1999} would correspond to a region where the departure is small somehow 10 percent or $\delta k \simeq \pi/10$. Outside the Fermi arc the surfaces are too far from each other to display a visible transition from occupied momentum states to unoccupied ones.

For a simple estimation of the transition line, we consider the case of absence of normal hopping ($|4t| \ll  |\Delta_{PG}|$ and  $n_{\epsilon} \simeq 0$) where the hole flips its spin at each  hopping.
As a consequence, the Fermi surface is centered around the antinodal points $(0,\pm \pi)$ or  $(\pm \pi,0)$ according to the direction of the symmetry breaking.
Using Eq.(\ref{D}) and Eq.(\ref{fPG}), in Fig.\ref{fig:4}, we show the region of doping and temperature at which the canonical free energy  in the pseudogap state $f^{PG}_{CN}$  is lower than the one in the normal state $f^{HF}_{CN}$.
The repulsive potential
reduces to the expression $U=U_S =4[U_{d1}-2U_{d2}n + U_{d3}(n^2-|n_{\overline{\epsilon}}|^2)]$ and is shown in the second graph.

The energy gap observed at the nodal point in \cite{doi:10.1073/pnas.1209471109}  corresponds to the difference between chemical potentials  $\mu^{PG}_0=d f^{PG}_{CN}/dn$ and $\mu^{HF}_0=d f^{HF}_{CN}/dn$.  For the parameter values of Fig.\ref{fig:3} and $t_{pd}=1eV$, we find at zero temperature the gap value
$\mu^{PG}_0 - \mu^{HF}_0 \sim 1 eV$ which is still larger than the experimental values of about $0.15 eV$. This higher value is attributed to the drastic assumption of  absence of normal hopping.

\subsection{Strange metal and Fermi liquid}

At high temperature the cuprates behave like an insulator with high resistivity and is referred to as a strange metal phase.
In the present  model description, the strange metal phase is  explained by a very narrow band energy close to a flat band and by the singular  density of states Eq.(\ref{dos}) around half-filling different from a Fermi liquid. As an indication, the Sommerfeld finite temperature expansion Eq.(\ref{HFT}) is  singular for $\mu=0$ showing its inaccuracy  concerning for instance the determination of the specific heat.
The weak perturbation  condition  in this expansion imposes a bound for the transition to a Fermi liquid by fixing the temperature to
$k_B T_{SM-FL} \sim \sqrt{\mu |t|}/20$. The prefactor $1/20$ has been chosen to adjust the transition line
plotted in  Fig.\ref{fig:3} in agreement with experimental data.
The detailed mechanism for the strange transport phenomena however still needs to be clarified but could be based on these unconventional properties.

\subsection{Antiferromagnetism}

The mean field approach may be used to address the antiferromagnetism, but appears not to be precise enough, starting directly from $\hat H$ \cite{Baeriswyl_2009}. For this reason, we use a second SW transformation to derive the associated $t-J $ model following a procedure done in \cite{condmat4020057}.
This phase is more accurately described in terms of  an effective Heisenberg Hamiltonian at half-filling:
\begin{eqnarray}\label{AF}
\hat H_{AF} =\sum_{\vc{l},i=x,y} J \hat {\bf S}_{\vc{l}}.
\hat {\bf S}_{\vc{l}+\vc{1}_i} \,,
\end{eqnarray}
where we define the on-site spin operators $\hat {\f S}_{\vc{l}}=\sum_{\sigma, \sigma'} \hat d^\dagger_{\vc{l},\sigma} \tau_{\sigma,\sigma'} \hat d_{\vc{l},\sigma'}/2$  where $\tau_{\sigma,\sigma'}$ are the Pauli matrices, and the superexchange spin coupling $J$  to be determined as follows.

Starting from Eq.(\ref{H}) at half-filling, we isolate the contribution proportional to the hopping term
$\hat n_{d,\vc{l},\overline{\sigma}}(1- \hat n_{d,\vc{l}\pm \vc{1}_i,\overline{\sigma}})
\hat d^\dagger_{\vc{l},\sigma} \hat d_{\vc{l} \pm \vc{1}_i,\sigma}
$
whose the coefficent $t_A= t_0 - U_{t1}/2$ corresponds to the effective virtual hopping in the antiferromagnetic state.
To obtain the difference of energy $U_A$, we subtract the energy associated to a doublon and an absence of hole minus the energy associated to single occupancy everywhere in order to obtain the expression Eq.(\ref{UA}).
Numerically, we calculate
$t_A=0.15$  and $U_{A}= 7.8 \simeq U_{d}$ using the values of Fig.\ref{fig:3} and deduce that
\begin{eqnarray}
J=\frac{ 4 t_A^2}{U_{A}} \sim 0.011 \, .
\end{eqnarray}
Using the value $t_{pd} \sim 1.8 eV$ keeping the other values unchanged in Fig.\ref{fig:3}, the estimation matches the experimental value $J \sim 0.1 eV$.
With the addition of a phenomenological transverse spin coupling between the $Cu O_2$ layers, previous works
\cite{PhysRevLett.84.4994,Govinf,SINGH1998201,Loktev} have shown how to  determine the phase diagram for the antiferromagnetism as well as the critical temperature in agreement with observations. Furthermore, we notice that for $U_{pd}=0$ and $U_d \gg \epsilon_p$,  we recover $J \simeq 4t_{1/2}^2/U_d$ as in  \cite{condmat4020057} but with $t_0$  replaced by $t_{1/2}$ .




\section{Conclusions}

We develop a mean field model in a basis of states that includes correlations between the orbitals $p$ and $d$.
It predicts both superconducting and pseudogap phases and provides some insights for the transition between
the strange metal and the Fermi liquid phases. We have merged both  well-established concepts of screening developed by Kohn and Luttinger \cite{KL} and the superexchange developed by Anderson \cite{Anderson} into one unified theory
which is a mere elegant use of the Schrieffer-Wolff technique applied to the three-band model to cuprates.
The resulting formalism has successfully provided many explanations for the behavior of the cuprates and therefore appears robust for applications in solid-state theory. It suggests that room temperature superconductivity is possible  for significant screening in superexchange interactions and a lower energy gap $\epsilon_p$ \cite{doi:10.1073/pnas.2207449119}.

Further experimental evidences - such as transport properties, specific heat measurements, charge density wave \cite{doi:10.1073/pnas.1910411116} or nematicity \cite{doi:10.1073/pnas.2206481119} - are still needed to fully validate the relevance of the present approach.
Also, some studies \cite{PhysRevLett.96.197001} suggest a magnetic order associated with the pseudogap phase that could be related to the order parameter Eq.(\ref{PG}).

{\bf Acknowledgements:}
PN is supported by the European Union project: QRC-4-ESP under grant agreement number 101129663 and
is grateful to Joseph Betouras, Xenophon Zotos, Ioannis Rousochatzakis, James Annett, and Todor Mishonov.
for fruitful discussions.

\vspace*{0.3cm}

\appendix

\section{The Schrieffer-Wolff transformation for superexchange}\label{A}

Without any unitary transformation on the three-band Hamiltonian, the HFB method does not work since the averages on the hopping terms vanish when the orbitals $p$ are empty.
The Hamiltonian Eq.(\ref{H1}) is decomposed into an interacting contribution $\hat H_0 $ and a perturbation  term $\hat V$ with the tight-binding parameter $t_{pd}$.

In absence of hopping corresponding the zeroth order in the expansion, we need to obey the stability criterion
$U_d n\leq \epsilon_p+2U_{pd}$ which provides an upper bound on the doping $n$.
Otherwise there are holes also  filling
the $p$ orbitals. A comparison with the experimental value shows that the presence of the repulsive term $U_{pd}$ usually neglected favours stability.
The  hopping term $t_{pd}$ increases this upper bound beyond half filling by  lowering further the ground state energy.

The combination of the HFB {\it ansatz} with the SW transformation (or the SWHFB method) generates  the superexchange mechanism responsible for the antiferromagnetic phase and which is also valid beyond  half-filling.
In order to eliminate the linear first order contribution and deal with the second order as the leading one, we define the SW unitary transformation $\hat S$  with the property that:
\begin{eqnarray}\label{SV}
\hat H'_T= e^{\hat S} \hat H_T e^{-\hat S}= \hat H_0 + [\hat S, \hat V]/2 + {\cal{O}}(t^3_{pd}) \, ,
\end{eqnarray}
which is possible if the operator fulfills the condition:
\begin{eqnarray}
 \hat V= [\hat H_0, \hat S] \, .
\end{eqnarray}
We deduce the solution:
\begin{eqnarray}
\hat S =\lim_{\eta \rightarrow 0}\frac{1}{i} \int_0^\infty dt\, e^{i\hat H_0 t}  \hat V  e^{-i\hat H_0 t} e^{-\eta t} \, ,
\end{eqnarray}
where $\eta$ is a convergence parameter.
Defining the vacuum $|0_p \rangle$ of the p-state in the resulting representation, we obtain using Eqs.(\ref{H2}) and (\ref{H3}) a Hamiltonian for the $d$ state only:
\begin{widetext}
\begin{eqnarray}\label{A4}
\hat H&=&\langle 0_p |\hat H'_T | 0_p \rangle
=\sum_{\vc{l}} U_d \hat n_{d,\vc{l},\uparrow}\hat n_{d,\vc{l},\downarrow}
\nonumber \\
&+&t_{pd}^2\lim_{\eta \rightarrow 0} \sum_{\vc{l},\vc{l'},\sigma,\pm,\pm'}\sum_{
i,i'=x,y}
\int_0^\infty dt \frac{e^{-\eta t}}{2i} \left[
 \hat d^\dagger_{\vc{l},\sigma}e^{-iU_d \hat n_{d,\vc{l},\overline{\sigma}}t}
\langle 0_p | \hat p_{\vc{l} \pm \vc{1}_i/2,\sigma} e^{-i \hat H_\sigma t} \hat p^\dagger_{\vc{l'} \pm' \vc{1}_{i'}/2,\sigma} | 0_p \rangle
\hat d_{\vc{l'},\sigma}+ c.c.\right] \, ,
\end{eqnarray}
where
\begin{eqnarray}\label{A5}
\hat H_\sigma&=& \!\! \!
\sum_{\vc{l},i=x,y}
 \hat n_{p,\vc{l} + \vc{1}_i/2,\sigma}\left[\epsilon_p + U_{pd} \sum_{\sigma'} (\hat n_{d,\vc{l},\sigma'}+ \hat n_{d,\vc{l}+ \vc{1}_i,\sigma'})\right]
+ t_{pp} \left(\hat p^\dagger_{\vc{l} +\vc{1}_{i}/2,\sigma} + \hat p^\dagger_{\vc{l} +\vc{1}_{\overline i}+\vc{1}_i/2,\sigma}\right)\left(\hat p_{\vc{l}+\vc{1}_{\overline i}/2,\sigma}+\hat p_{\vc{l}+ \vc{1}_i+\vc{1}_{\overline i}/2,\sigma}\right)  \, .
\nonumber \\
\end{eqnarray}

\end{widetext}
The term $U_p$ disappears since up to the second order, only one virtual hole occupies the $p$ orbitals without any interaction with the others. The  Hamiltonian $\hat H_\sigma$ describes the dynamics of virtual hole excitation  within the lattice associated to the $p$ orbitals in a disordered environment caused by the repulsive potential of the other holes in the $d$ orbitals. The resulting virtual process  generates  an effective hopping of holes between the $d$ sites. From this
unitary transformation the initial quantum state  displaying inter-particle correlation is transformed into  a state that can be dealt with the HFB  mean field approach involving an effective attraction between holes and therefore superconductivity.

However, the expressions Eq.(\ref{A4}) and Eq.(\ref{A5}) are still too complicated to be tractable. Another simplification is to neglect the hopping between the $p$ orbitals by setting $t_{pp}=0$. In these conditions, the eigenstates of $\hat H_\sigma$ are known exactly and correspond to an insulator.
While this approximation is not entirely realistic, it serves as a proof of principle for pairing. After averaging over the vacuum of the $p$ state,
it results into the transformed Hamiltonian:
\begin{widetext}
\begin{eqnarray}
\hat H&=&
-\frac{t_{pd}^2}{2} \!\!\!
\sum_{\vc{l},\sigma,\pm,i=x,y}  \left[ \frac{\hat n_{d,\vc{l},\sigma} }
{\epsilon_p -U_d^* \hat n_{d,
\vc{l},\overline{\sigma}}+U_{pd}\sum_{\sigma'}\hat n_{d,
\vc{l}  \pm  \vc{1}_{i},\sigma'}}+\frac{\hat d^\dagger_{\vc{l},\sigma} \hat d_{\vc{l}\pm \vc{1}_i,\sigma}}
{\epsilon_p -U_d^* \hat n_{d,
\vc{l},\overline{\sigma}}+U_{pd}\hat n_{d,
\vc{l}  \pm  \vc{1}_{i},\overline{\sigma}}}+c.c. \right]
+ \sum_{\vc{l}} U_{d} \hat n_{d,\vc{l},\uparrow}\hat n_{d,\vc{l},\downarrow} \, .
\nonumber \\
\end{eqnarray}
To determine the interaction potential, we use the relation:
\begin{eqnarray}
 &&\frac{1 }
{\epsilon_p -U_d^* \hat n_{d,
\vc{l},\overline{\sigma}}+U_{pd}\sum_{\sigma'}\hat n_{d,
\vc{l}  \pm  \vc{1}_{i},\sigma'}}=
\nonumber \\ 
&&(1-\hat n_{d,
\vc{l},\overline{\sigma}}) \left[
\frac{(1-\hat n_{d,
\vc{l}\pm  \vc{1}_{i},\sigma})(1-\hat n_{d,
\vc{l}\pm  \vc{1}_{i},\overline{\sigma}})}
{\epsilon_p}
+
\frac{ \sum_{\sigma'}\hat n_{d,
\vc{l}\pm  \vc{1}_{i},\sigma'}(1-\hat n_{d,
\vc{l}\pm  \vc{1}_{i},\overline{\sigma}'})}
{\epsilon_p +U_{pd}}+
\frac{ \hat n_{d,
\vc{l}\pm  \vc{1}_{i},\sigma}\hat n_{d,
\vc{l}\pm  \vc{1}_{i},\overline{\sigma}}}
{\epsilon_p +2U_{pd}}\right]+
\nonumber \\
&&\hat n_{d,
\vc{l},\overline{\sigma}}
\left[
\frac{(1-\hat n_{d,
\vc{l}\pm  \vc{1}_{i},\sigma})(1-\hat n_{d,
\vc{l}\pm  \vc{1}_{i},\overline{\sigma}})}
{\epsilon_p -U_d^*}
+
\frac{ \sum_{\sigma'}\hat n_{d,
\vc{l} \pm  \vc{1}_{i},\sigma'}(1-\hat n_{d,
\vc{l} \pm  \vc{1}_{i},\overline{\sigma}'})}
{\epsilon_p -U_d^*+U_{pd}}+
\frac{ \hat n_{d,
\vc{l}\pm  \vc{1}_{i},\sigma}\hat n_{d,
\vc{l}\pm  \vc{1}_{i},\overline{\sigma}}}
{\epsilon_p -U_d^*+2U_{pd}}\right] \, .
\end{eqnarray}
We obtain:
\begin{eqnarray} \label{H}
\hat H
&= &
-\!\!\!\!\!\!\!\sum_{\vc{l},\sigma,\pm,i=x,y}  \left[\frac{[t_0 -\hat n_{d,
\vc{l},\overline{\sigma}}(U_{t1}- U_{t2}\hat n_{d,
\vc{l}\pm\vc{1}_i,\overline{\sigma}})] \hat d^\dagger_{\vc{l},\sigma}\hat d_{\vc{l} \pm\vc{1}_i,\sigma} +c.c.}{2}+ [t_0-(U_{d1}- U_{d2}\hat n_{d,
\vc{l},\overline{\sigma}})\sum_{\sigma'} \hat n_{d,
\vc{l}\pm\vc{1}_i,\sigma'}]\hat n_{d,
\vc{l},\sigma}\right]
\nonumber \\
&+& \sum_{\vc{l}} \left[U_{dR}+U_{d3} \sum_{\pm,i=x,y}\hat n_{d,
\vc{l}\pm\vc{1}_i,\uparrow}\hat n_{d,
\vc{l}\pm\vc{1}_i,\downarrow} \right]\hat n_{d,\vc{l},\uparrow}\hat n_{d,\vc{l},\downarrow} \, ,
\end{eqnarray}
with the positive parameters (for the physical values considered here):
\begin{eqnarray}\label{U}
t_0 &= & \frac{t_{pd}^2}{\epsilon_p} \, , \quad\quad U_{t1}=
t_{pd}^2\left(\frac{2}{\epsilon_p}-\frac{1}{\epsilon_p-U_d^* } -\frac{1}{\epsilon_p+ U_{pd} }\right) \, ,\quad \quad
U_{t2}=t_{pd}^2\left(\frac{1}{\epsilon_p}-\frac{1}{U_{pd} +\epsilon_p} 
+\frac{1}{U_d^* -\epsilon_p}-\frac{1}{U_d^* -U_{pd} -\epsilon_p}\right) 
\, ,
\nonumber \\
U_{dR}
&=&U_d +\frac{8t_{pd}^2}{\epsilon_p} +\frac{8t_{pd}^2}{(U_d^* -\epsilon_p)} \, ,
\quad \quad 
U_{d1}=t_{pd}^2 \left(\frac{1}{\epsilon_p}-\frac{1}{U_{pd} +\epsilon_p} \right) \, , \quad \quad  U_{d2}=U_{t2} +\frac{t_{pd}^2 U_{pd}^2}{\epsilon_p(\epsilon_p+ U_{pd})(\epsilon_p+ 2U_{pd})} \, ,
\nonumber \\
U_{d3}&=&4U_{pd}^2t_{pd}^2\left[\frac{1}{\epsilon_p(\epsilon_p+U_{pd})(\epsilon_p+2U_{pd})}+\frac{1}{(U_d^*-\epsilon_p)(U_d^*-\epsilon_p-U_{pd})(U_d^*-\epsilon_p-2U_{pd})}\right] \, ,
\end{eqnarray}
where  $U_d^* =U_d - U_{pd}$.
When  the potential $U_{pd}$ is neglected in Eq.(\ref{H}), we find a Hamiltonian similar to the Hubbard model with the only difference of the additional quartic term  of the form $\hat n_{d,\vc{l},\overline{\sigma}}\hat d^\dagger_{\vc{l},\sigma} \hat d_{\vc{l} \pm \vc{1}_i,\sigma}$. Around half filling,  this term  reduces to the simple hopping
by setting $\langle \hat n_{d,\vc{l},\overline{\sigma}}\rangle =1/2$.
Up to some constant term, we recover by this procedure the Hubbard Hamiltonian:
\begin{eqnarray} \label{Hub}
\hat H_{Hub}
&= &
-\!\!\!\!\!\!\!\sum_{\vc{l},\sigma,\pm,i=x,y} t_{1/2}
\hat d^\dagger_{\vc{l},\sigma}\hat d_{\vc{l} \pm\vc{1}_i,\sigma}
+\sum_{\vc{l}} U_{dR}\hat n_{d,
\vc{l},\uparrow}\hat n_{d,
\vc{l},\downarrow} \, ,
\end{eqnarray}
where $t_{1/2}=t_0-U_{t1}/2 \simeq t_0/2$. This derivation has also been achieved in \cite {ZR, PhysRevB.44.7504,PhysRevB.71.134527} using different methods but by treating the influence of $U_{pd}$ only at the mean field level.

\section{Mean field energy and gap equations}\label{B}

In the new representation, we choose a basis of state such that $\langle \hat d^\dagger_{\vc{l},\sigma}\hat d_{\vc{l},\overline{\sigma}}\rangle=0$.
We use the Wick theorem with the assumption that only d-wave nonlocal pairing with terms of the form  $\langle \hat d_{\vc{l},\sigma} \hat d_{\vc{l}+ \vc{1}_x,\overline \sigma}\rangle= -\langle \hat d_{\vc{l},\sigma} \hat d_{\vc{l}+ \vc{1}_y,\overline \sigma}\rangle$  is relevant excluding local s-wave pairing $\langle \hat d_{\vc{l},\sigma} \hat d_{\vc{l},\overline{\sigma}}\rangle=0$. The ansatz contains only the four averages $n$,$n_\epsilon$,$n_{\overline{\epsilon}}$ and $m$. Since we aim at an easily tractable model, other averages like the tripled Cooper pairs \cite{Annett1999} or other nearest neighbour averages have been discarded.
Using the assumptions $\langle
\hat d_{\vc{l},\downarrow}\hat d_{\vc{l}+ \vc{1}_{i},\uparrow}
\rangle= \langle 
\hat d_{\vc{l},\downarrow}\hat d_{\vc{l} -\vc{1}_{i},\uparrow}
\rangle=\langle \hat d^\dagger_{\vc{l}\pm \vc{1}_{i},\uparrow}\hat d^\dagger_{\vc{l},\downarrow}
\rangle$, we obtain terms with the form:
\begin{eqnarray} 
&&\sum_{\vc{l},\pm,i=x,y} \frac{
\langle \hat d^\dagger_{\vc{l}\pm \vc{1}_{i},\uparrow}\hat d^\dagger_{\vc{l},\downarrow}
\rangle
\langle 
\hat d_{\vc{l},\downarrow}\hat d_{\vc{l}\pm \vc{1}_{i},\uparrow}
\rangle}{N}=
\sum_{\vc{l},\pm} \frac{
\langle (\hat d^\dagger_{\vc{l}\pm \vc{1}_{x},\uparrow}-
\hat d^\dagger_{\vc{l}\pm \vc{1}_{y},\uparrow})\hat d^\dagger_{\vc{l},\downarrow}
\rangle 
\langle 
\hat d_{\vc{l},\downarrow}(\hat d_{\vc{l}\pm \vc{1}_{x},\uparrow}
 -
 \hat d_{\vc{l}\pm \vc{1}_{y},\uparrow})
\rangle }{2N} 
\nonumber \\
&=&\sum_{\vc{l}} \frac{\sum_{\pm}
\langle (\hat d^\dagger_{\vc{l}\pm \vc{1}_{x},\uparrow}-
\hat d^\dagger_{\vc{l}\pm \vc{1}_{y},\uparrow})\hat d^\dagger_{\vc{l},\downarrow}
\rangle 
\sum_{\pm}
\langle 
\hat d_{\vc{l},\downarrow}(\hat d_{\vc{l}\pm \vc{1}_{x},\uparrow}
 -\hat d_{\vc{l}\pm \vc{1}_{y},\uparrow})
\rangle }{4N}=4|m|^2 \, .
\end{eqnarray}
From the Wick decomposition,
we determine the averages:
\begin{eqnarray}
\sum_{\vc{l}} \frac{\langle \hat n_{d,\vc{l},\uparrow} \hat n_{d,\vc{l},\downarrow} \rangle}{N} &=&n^2 \, ,
\\
\sum_{\vc{l},\sigma,\sigma',\pm,i=x,y} \frac{\langle \hat n_{d,
\vc{l},\sigma} \hat n_{d,
\vc{l}\pm\vc{1}_i,\sigma'}\rangle}{N}
&=&8[2n^2-n_\epsilon^2- |n_{\overline{\epsilon}}|^2+|m|^2] \, ,
\\
\sum_{\vc{l},\sigma,\sigma',\pm,i=x,y} \frac{\langle \hat n_{d,
\vc{l},\sigma}\hat n_{d,
\vc{l},\overline{\sigma}} \hat n_{d,
\vc{l}\pm\vc{1}_i,\sigma'}\rangle}{N}
&=&16n[n^2  -n_\epsilon^2- |n_{\overline{\epsilon}}|^2+|m|^2] \, ,
\\
\sum_{\vc{l},\sigma,\pm,i=x,y} \frac{\langle \hat n_{d,
\vc{l},\overline{\sigma}} \hat n_{d,
\vc{l}\pm\vc{1}_i,\overline{\sigma}}\hat d^\dagger_{\vc{l},\sigma}\hat d_{\vc{l} \pm\vc{1}_i,\sigma}\rangle}{N}
&=&8n_{\epsilon}[n^2-n_\epsilon^2+|n_{\overline{\epsilon}}|^2-|m|^2] \, ,
\\
\sum_{\vc{l},\sigma,\pm,i=x,y} \!\!\!\!\!\frac{\langle \hat n_{d,
\vc{l},\overline{\sigma}} \hat d^\dagger_{\vc{l},\sigma}\hat d_{\vc{l} \pm\vc{1}_i,\sigma}\rangle}{N}
&=&8n_{\epsilon}n  \, ,
\\
\!\!\!\!\!
\sum_{\vc{l},\pm,i=x,y}\!\!\!\!\!\frac{\langle \hat n_{d,
\vc{l},\uparrow}\hat n_{d,
\vc{l},\downarrow} \hat n_{d,
\vc{l}\pm\vc{1}_i,\uparrow}\hat n_{d,
\vc{l}\pm\vc{1}_i,\downarrow}\rangle}{N}
&=&4[n^4  -2n^2 n_\epsilon^2 +2 n^2(|m|^2- |n_{\overline{\epsilon}}|^2)
\nonumber \\
&+&n_\epsilon^4+(|n_{\overline{\epsilon}}|^2-|m|^2)^2 
+2n^2_\epsilon(|m|^2-|n_{\overline{\epsilon}}|^2)]  \, .
\end{eqnarray}
The grand canonical free energy per site and spin is function of four minimization parameters:
\begin{eqnarray}
&&f(n,n_\epsilon,m,m^*,n_{\overline{\epsilon}},n^*_{\overline{\epsilon}})= \frac{\langle \hat H -\mu_0 \sum_{\vc{l},\sigma}
\hat n_{d,\vc{l},\sigma} \rangle}{2N}  \, ,
\nonumber \\
&=&-4t_0 n_\epsilon-(\mu_0+4t_0)n+4U_{t1}n_\epsilon n+
\frac{U_{dR}}{2}n^2+4U_{d1}[2n^2-n_\epsilon^2- |n_{\overline{\epsilon}}|^2+|m|^2]
\nonumber \\
&-&8U_{d2}n[n^2   -n_\epsilon^2- |n_{\overline{\epsilon}}|^2+|m|^2]
-4U_{t2} n_{\epsilon}[n^2 -n_\epsilon^2+|n_{\overline{\epsilon}}|^2-|m|^2]
\nonumber \\
&+&2U_{d3}[n^4 -2n^2 n_\epsilon^2 +2(n^2-n^2_\epsilon )(|m|^2- |n_{\overline{\epsilon}}|^2)
+n_\epsilon^4+(|n_{\overline{\epsilon}}|^2-|m|^2)^2] \, ,
\end{eqnarray}
where $\mu_0$ is the chemical potential.
In the case the term $1/\epsilon_p$ is dominant in the Hamiltonian, the contribution in the first order in $t_{pp}$ in Eqs.(\ref{A4}) and (\ref{A5}) becomes significant and leads to an effective renormalization:
\begin{eqnarray}\label{tppeff}
t_0 \rightarrow t_0 - \frac{4 t^2_{pd} t_{pp}^{eff} }{\epsilon_p^2}
\, ,
\quad \quad \quad
U_{t1} \rightarrow U_{t1}   - \frac{4 t^2_{pd} t_{pp}^{eff}}{\epsilon_p^2} \, ,
\end{eqnarray}
The hopping term $t_{pp}$ has been replaced by an effective hopping $t_{pp}^{eff}$ determined after resumming all terms in the expansion of the bare parameter $t_{pp}/\epsilon_p$.

We define the effective chemical potential, the effective hopping and gap functions respectively as:
\begin{eqnarray}\label{mu}
\mu&=&-\frac{\partial f}{\partial n}=\mu_0+4t_0
-(4U_{t1}-8U_{t2}n) n_\epsilon -(U_{dR}+16U_{d1})n
 \nonumber \\
&+&8U_{d2}(3 n^2   -n_\epsilon^2- |n_{\overline{\epsilon}}|^2+|m|^2)
-8U_{d3}[n^3 -n (n_\epsilon^2 -|m|^2 + |n_{\overline{\epsilon}}|^2)] \, ,
\\ \label{t}
t&=&-\frac{1}{4}\frac{\partial f}{\partial n_\epsilon}
=t_0-U_{t1}n +U_{t2}
(n^2 -3n_\epsilon^2+|n_{\overline{\epsilon}}|^2-|m|^2)+[2U_{d1}-4U_{d2}n+ 2U_{d3}(n^2-n_\epsilon^2-|n_{\overline{\epsilon}}|^2+|m|^2)]n_\epsilon  \, ,
\nonumber \\ \\ \label{D}
\Delta&=&-\frac{\partial f}{\partial m^*}=-Um  \, ,
\quad \quad \quad \quad  U=4[U_{d1}-2U_{d2}n +U_{t2}n_\epsilon+U_{d3}(n^2
+n_\epsilon^2-|n_{\overline{\epsilon}}|^2+|m|^2)] \, ,
\\ \label{DPG}
\Delta_{PG}&=&-\frac{\partial f}{\partial n^*_{\overline{\epsilon}}} = Un_{\overline{\epsilon}} \, ,
\end{eqnarray}
where we  use the partial derivative notation
$\partial/\partial n^*_{\overline{\epsilon}}=[\partial/\partial ({\rm Re}n_{\overline{\epsilon}}) +i\partial/\partial ({\rm Im} n_{\overline{\epsilon}})]/2$. 

We need to determine the gap equations for these Lagrange parameters.
For that purpose, we define the Fourier transform:
\begin{eqnarray}
\hat d_{\vc{k},\sigma}=\sum_{\vc{l}} \frac{e^{-i\vc{k}.\vc{l}}}{\sqrt{N}}
\hat d_{\vc{l},\sigma} \, ,
\quad \quad 
\hat d_{\vc{l},\sigma}=\sum_{\vc{k}} \frac{e^{i\vc{k}.\vc{l}}}{\sqrt{N}}
\hat d_{\vc{k},\sigma}  \, ,
\end{eqnarray}
and we use the effective mean field  Hamiltonian:
\begin{eqnarray}\label{mf}
\hat H_{MF} &=&  \sum_{\vc{k}}\left(\sum_{\sigma}
(\epsilon_{\vc{k}}-\mu) \hat n_{d,\vc{k},\sigma}-
\Delta_{\vc{k},\sigma}  
\hat d^\dagger_{\vc{k},\sigma}\hat d_{\vc{k},\overline{\sigma}}\right) 
-\Delta^*_\vc{k}\hat d_{\vc{k},\downarrow}\hat d_{-\vc{k},\uparrow} - \Delta_\vc{k}\hat d^\dagger_{-\vc{k},\uparrow}\hat d^\dagger_{\vc{k},\downarrow}  \, ,
\end{eqnarray}
where we define  $\epsilon_{\vc{k}}=-2t(\cos k_x + \cos k_y )$, $\Delta_{\vc{k},\uparrow}=\Delta^*_{\vc{k},\downarrow}= (\cos k_x - \cos k_y)\Delta_{PG} /2$  and $\Delta_\vc{k}=(\cos k_x - \cos k_y)\Delta/2$. We note the antisymmetry $\epsilon_{\vc{k}\pm \vc{Q}}=- \epsilon_{\vc{k}}>0$ where $\vc{Q}=(\pi,\pi)$ is the bulk supermodulation vector.

The Hamiltonian (\ref{mf}) is quadratic in the fermion operators and therefore can be diagonalized by means of successive transformations. The first is:
\begin{eqnarray}
\hat d_{\vc{k},\sigma}= \cos(\beta_{\vc{k}})\hat b_{\vc{k},\sigma}   + \sigma \sin(\beta_{\vc{k}})\hat b^\dagger_{-\vc{k},\overline{\sigma}}
\, .
\end{eqnarray}
The Hamiltonian becomes:
\begin{eqnarray}
\hat H_{MF} &=&  \sum_{\vc{k}, \sigma}
E_{\vc{k}} (\hat n_{b,\vc{k},\sigma}-\frac{1}{2})-
\Delta_{\vc{k},\sigma} 
\hat b^\dagger_{\vc{k},\sigma}\hat b_{\vc{k},\overline{\sigma}}
-\mu/2 \, ,
\end{eqnarray}
where $E_{\vc{k}}=\sqrt{(\epsilon_{\vc{k}}- \mu)^2 + |\Delta_\vc{k}|^2}$.
The second transformation is:
\begin{eqnarray}
\hat b_{\vc{k},\sigma}=\frac{1}{\sqrt{2}} \left(\hat c_{\vc{k},\sigma}   + \frac{\overline{\sigma}\Delta_{\vc{k},\sigma}}{|\Delta_{\vc{k},\sigma}|}\hat c_{\vc{k},\overline{\sigma}} \right)  \, ,
\end{eqnarray}
to obtain the diagonal form:
\begin{eqnarray}\label{mfc'}
\hat H_{MF} &=&  \sum_{\vc{k}, \sigma}
(E_{\vc{k}} +\sigma |\Delta_{\vc{k},\sigma}| )  (\hat n_{c,\vc{k},\sigma}-\frac{1}{2})-\mu/2  \, .
\end{eqnarray}
We assume a 
Fermi-Dirac distribution $\langle \hat n_{c,\vc{k},\sigma}\rangle =1/[\exp[\beta (E_{\vc{k}} +\sigma |\Delta_{\vc{k},\sigma}| )]+1]$ where $\beta=1/k_B T$ is the inverse temperature.
By taking the derivatives using (\ref{mf}) and (\ref{mfc'}) and  the definitions of the order parameters, we deduce the relations:
\begin{eqnarray}\label{ng}
n=-\frac{1}{2N}\langle\frac{\partial \hat H_{MF} }{\partial \mu}\rangle =\frac{1}{2}+\frac{1}{2N}\sum_{\vc{k},\sigma }
\frac{( \mu - \epsilon_{\vc{k}})}{2E_{\vc{k}}}
\tanh\left(\frac{\beta(E_{\vc{k}} +\sigma |\Delta_{\vc{k},\sigma}|)}{2}\right) \, ,
\end{eqnarray}
\begin{eqnarray}\label{neg}
n_\epsilon=-\frac{1}{8N}\langle\frac{\partial \hat H_{MF} }{\partial t}\rangle =\frac{1}{2N}\sum_{\vc{k},\sigma }
\frac{(\cos k_x + \cos k_y)( \mu - \epsilon_{\vc{k}})}{4E_{\vc{k}}}
\tanh\left(\frac{\beta(E_{\vc{k}} +\sigma |\Delta_{\vc{k},\sigma}|)}{2}\right) \, ,
\end{eqnarray}
\begin{eqnarray}\label{m}
m=-\langle \frac{\partial }{\partial \Delta^*}\frac{\hat H_{MF} }{2N}   \rangle
=\frac{1}{2N}\sum_{\vc{k},\sigma}
\frac{(\cos k_x - \cos k_y)^2\Delta}{8E_{\vc{k}}}
\tanh\left(\frac{\beta(E_{\vc{k}} +\sigma |\Delta_{\vc{k},\sigma}|)}{2}\right)  \, ,
\end{eqnarray}
\begin{eqnarray}\label{nPG}
n_{\overline{\epsilon}}=-\langle
\frac{\partial }{\partial \Delta^*_{PG}} \frac{\hat H_{MF} }{2N}
\rangle=\frac{1}{2N}\sum_{\vc{k},\sigma}
\frac{\sigma(\cos k_x - \cos k_y)^2\Delta_{PG}}{8|\Delta_{\vc{k},\sigma}|}\tanh\left(\frac{\beta(E_{\vc{k}} +\sigma |\Delta_{\vc{k},\sigma}|)}{2}\right) \, .
\end{eqnarray}
Together with Eqs.(\ref{mu}),(\ref{t}),(\ref{D}) and  (\ref{DPG}), the Eqs.(\ref{ng}),(\ref{neg}),(\ref{m}) and (\ref{nPG}) form a set of generalized gap equations to be solved.

We need also the explicit expression for the canonical free energy per hole and per spin corresponding to the average Hamiltonian minus the entropy times the temperature:
\begin{eqnarray}\label{CN}
f_{CN}= f+ \mu n -\frac{1}{2N} \sum_\vc{k, \sigma}\left[
\frac{(E_{\vc{k}} +\sigma |\Delta_{\vc{k},\sigma}|)}{\exp\left[\beta (E_{\vc{k}} +\sigma |\Delta_{\vc{k},\sigma}|)\right] +1} -
\ln\left(1+\exp[-\beta (E_{\vc{k}} +\sigma |\Delta_{\vc{k},\sigma}|)]\right)/\beta\right]  \, .
\end{eqnarray}
This quantity shall be used in appendix \ref{C} to compare the PG state with the  HF state to determine the phase from the lowest energy.

\section{Analytical estimations}\label{C}

\subsection{Hartree-Fock approximation for the density   state}

In this subsection, we carry out analytical calculations for a square lattice to determine the density of state, the particle density and the hopping energy as a fucntion of the effective chemical potential $\mu$. Using the elliptic functions,
the density of state per spin is defined as:
\begin{eqnarray}\label{dos}
D(\epsilon)&=&\frac{dn}{d\mu}(\mu=\epsilon)= \sum_{\vc{k}} \frac{\delta(\epsilon-\epsilon_\vc{k})}{N}=
\frac{4}{\pi^2}\int_0^1  \int_0^{1-u} \!\!\!\! \frac{du dv \delta(|\epsilon|-4|t|u)}{\sqrt{[1-(u+v)^2][1-(u-v)^2]}} \, ,\quad \quad u, v=(\cos(k_x) \pm \cos(k_y))/2
\nonumber \\
&=& \frac{4}{\pi^2}\int_0^1 \frac{du \,  \delta(|\epsilon|-4|t|u)}{1+u}\int_0^{\pi/2}
\frac{d\theta}{\sqrt{1- \left(\frac{1-u}{1+u}\sin\theta\right)^2}} \, , \quad
\quad \quad
v=(1-u)\sin\theta  \, ,
\nonumber \\
&=& 
\frac{1^+(4t-|\epsilon|)}{2\pi^2 |t|}K(\sqrt{1-(\epsilon/4t)^2})
\, ,
\quad \quad \quad
K(k)=\int_0^{\pi/2} \frac{d\theta}{\sqrt{1-k^2 \sin^2 \theta}} \, ,
\end{eqnarray}
where we use the identity:
\begin{eqnarray}
K(k)= \frac{2}{1+\sqrt{1-k^2}}K\left(\frac{1-\sqrt{1-k^2}}{1+\sqrt{1-k^2}}\right) \, .
\end{eqnarray}
Around half-filling,
the  density has the simpler expression $D(\epsilon)\cong\ln|16t/\epsilon|/(2\pi^2 |t|)$ showing that the density of state becomes infinite at $\mu=0$. The error bar for this approximation is about $10\%$ in the worst case.
In absence of pairing, we find the mean field Hartree-Fock results at zero temperature developed around half-filling :
\begin{eqnarray}\label{n}
n^{HF}&=&\int_{-4|t|}^\mu D(\epsilon)d\epsilon \cong 1/2+\frac{\mu}{2\pi^2|t|}[1+\ln|16t/\mu|]  \, ,
\\ \label{ne}
n^{HF}_\epsilon&=&-\int_{-4|t|}^\mu \frac{\epsilon}{4|t|} D(\epsilon)d\epsilon
=-\frac{2}{\pi^2}\left[\left(\frac{\mu}{4t}\right)^2 K\left(\sqrt{1-(\mu/4t)^2}\right) - E\left(\sqrt{1-(\mu/4t)^2}\right) \right] 
\nonumber \\
&\cong &
\frac{2}{\pi^2} -\frac{1}{\pi^2}\left(\frac{\mu}{4t}\right)^2 \ln|16t\sqrt{e}/\mu|  \, ,
\end{eqnarray}
where $E(k)=\int_0^{\pi/2} d\theta \sqrt{1-k^2 \sin^2 \theta}\cong 1 -\frac{1-k^2}{4}\ln((1-k^2)e/16)$ in which we use the neperian number $e$.
At finite temperature and $\mu \not=0$, we use the Sommerfeld expansion:
\begin{eqnarray}
\int _{-\infty }^{\infty }{\frac {H(\epsilon )}{e^{\beta (\epsilon -\mu )}+1}}\,\mathrm {d} \epsilon =\int _{-\infty }^{\mu }H(\epsilon )\,\mathrm {d} \epsilon +{\frac {\pi ^{2}}{6}}\left({\frac {1}{\beta }}\right)^{2}H^{\prime }(\mu )+O\left({\frac {1}{\beta \mu }}\right)^{4}\, ,
\end{eqnarray}
to obtain the leading order corrections
\begin{eqnarray}\label{HFT}
n^{HFT} - n^{HF}\cong  \frac {\pi ^{2}}{6}\left({\frac {1}{\beta }}\right)^{2}\frac{\partial D(\mu )}{\partial \mu} \simeq - \frac {1}{12\beta^2 \mu |t| }  \, ,
\quad \quad
n_\epsilon^{HFT} -n_\epsilon^{HF} \cong -\frac {\pi ^{2}}{24 |t|}\left({\frac {1}{\beta }}\right)^{2}\frac{\partial (\mu D(\mu ))}{\partial \mu} \simeq
\frac {\ln|e\mu/16 t|}{48\beta^2 t^2 }  \, .
\end{eqnarray}
A small correction corresponds to a Fermi liquid while a large correction corresponds to a strange metal. Using the notation $u, v$
 and $\theta$ in (\ref{dos}), the density of state for pairing is defined as:
\begin{eqnarray}\label{Dd}
D_d(\epsilon)&=&
\sum_{\vc{k}} \frac{v^2 \delta(\epsilon-\epsilon_\vc{k})}{N}=
\frac{4}{\pi^2}\int_0^1  \int_0^{1-u} \!\!\!\! \frac{du dv\, v^2\delta(|\epsilon|-4|t|u)}{\sqrt{[1-(u+v)^2][1-(u-v)^2]}}
\nonumber \\
&=& \frac{4}{\pi^2}\int_0^1 \frac{du \,  \delta(|\epsilon|-4|t|u)}{1+u}\int_0^{\pi/2}
\frac{d\theta (1-u)^2 \sin^2\theta}{\sqrt{1- \left(\frac{1-u}{1+u}\sin\theta\right)^2}}
\nonumber \\
&=& 
\frac{1}{\pi^2|t|}\left[(
1+(\epsilon/4t)^2)K(\sqrt{1-(\epsilon/4t)^2})/2-E(\sqrt{1-(\epsilon/4t)^2})\right]
\cong \frac{1}{2\pi^2|t|}\ln|16t/(e^2\epsilon)| \, ,
\end{eqnarray}
where we use:
\begin{eqnarray}
E(k)= (1+\sqrt{1-k^2})E\left(\frac{1-\sqrt{1-k^2}}{1+\sqrt{1-k^2}}\right) -\sqrt{1-k^2}K(k) \,.
\end{eqnarray}

\subsection{Conditions for pairing}

Using the gap equation defined from the combination of Eq.(\ref{D}) and Eq.(\ref{m}), we determine the transition line for superconductivity.
For low temperature and for vanishing gap for a d-wave, we write successively using Eq.(\ref{Dd}):
\begin{eqnarray}
\frac{1}{|U|}=\frac{m}{\Delta}&=&
\frac{1}{N}\sum_{\vc{k}}\frac{(\Delta_ 
\vc{k}/\Delta)^2\tanh(\beta(\epsilon_{\vc{k}}-\mu)/2)}{2(\epsilon_{\vc{k}}-\mu) }
=\int_{-4|t|}^{4|t|} \!\!\!d\epsilon\,D_d(\epsilon)\frac{\tanh(\beta(\epsilon-\mu)/2)}{2(\epsilon-\mu)}
\nonumber \\
&\stackrel{|\beta \mu| \gg 1}{=}&\int_{-4|t|}^{4|t|} \!\!\!d\epsilon\,
D_d(\mu ) \frac{\tanh(\beta(\epsilon-\mu)/2)}{2(\epsilon-\mu)}
+ F(\mu)
\nonumber \\
&=&\frac{D_d(\mu )}{2} \left[ \ln(x) \tanh(x)\bigg |_{-\beta(4|t| +\mu)/2}^{\beta(4|t|-\mu)/2}
- \int_{-\beta(4|t| +\mu)/2}^{\beta(4|t|-\mu)/2} dx \frac{\ln(x)}{\cosh^2(x)} \right]+ F(\mu)
\nonumber \\
&\stackrel{|\beta \mu|\gg 1 }{=}&  D_d(\mu )(\ln( \beta \sqrt{4t^2-\mu^2/4})+C_1 )
+ F(\mu)
\end{eqnarray}
We use the Euler-Mascheroni constant ${\bf C}= 0.577\dots$ in the integral
\begin{eqnarray}
C_1=-\int_0^\infty dx \frac{\ln(x)}{\cosh^2(x)}=\ln(\frac{4 e^{\bf C}}{\pi})=0.8188  \,,
\end{eqnarray}
and the function
\begin{eqnarray}
F(\mu)= \int_{-4|t|}^{4|t|} \!\!\!d\epsilon\,
\frac{D_d(\mu)-D_d(\epsilon)}{2|\epsilon-\mu|}
\stackrel{\mu \ll 4|t|  }{=} \int_{-4|t|}^{4|t|} \!\!\!d\epsilon\,
\frac{\ln|\epsilon/\mu|}{4\pi^2 |t||\epsilon-\mu|}
=\frac{1}{4\pi^2 |t|}
\left[\ln^2(4|t|/\mu) -\frac{1}{2} {\rm Li_2}\left( (\mu/4t)^2 \right) -\zeta(2) \right]  \,,
\end{eqnarray}
that involves the dilogarithm function:
\begin{eqnarray}
{\rm Li_2}\left(x \right)= \sum_{k=1}^\infty \frac{x^k}{k^2} \quad \quad \quad \zeta(2)={\rm Li_2}\left(1 \right)=\pi^2/6 \,.
\end{eqnarray}
We deduce the  critical temperature
\begin{eqnarray}\label{Tc}
k_B T_{cs}&=&
\sqrt{4t^2-\mu^2/4} \exp\left[C_1 -
(1/|U| - F(\mu))/D_d(\mu)\right]
\\ \label{Tca}
&\stackrel{\mu \ll 4|t|  }{=}& \sqrt{4t^2-\mu^2/4} \exp\left[C_1 -\frac{2\pi^2|t|}{|U|\ln|16 t /(\mu e^2) |} -\frac{\ln^2(4|t|/\mu) -\frac{1}{2} {\rm Li_2}\left( (\mu/4t)^2 \right) -\pi^2/6}{2 \ln|16 t/(\mu e^2)|}\right]  \,.
\end{eqnarray}
This last formula is valid for $\beta \mu \gg 1$. Close to half-filling, the error bar for the analytical formula Eq.(\ref{Tca}) is about $10\%$ in comparison to the numerical expression Eq.(\ref{Tc}).

\subsection{The frustrated hole regime or pseudogap}

In order to obtain analytical results, we study the PG regime where the normal hopping is negligible  $|t| \ll |\Delta_{PG}|=|U_S n_{\overline{\epsilon}}|$.
The mean field Hamiltonian (\ref{mfc'}) becomes:
\begin{eqnarray}\label{mfpg}
\hat H_{MF} &=&  \sum_{\vc{k}, \sigma}
\left(\sigma  |\Delta_{PG}| \frac{\cos k_x - \cos k_y}{2} - \mu \right)  \hat n_{c,\vc{k},\sigma}  \, .
\end{eqnarray}
With the help of the techniques developed in Eqs.(\ref{dos}),(\ref{n}),(\ref{ne}) and the notation $\mu_S=4\mu t/|\Delta_{GP}|$,  we determine Eqs.(\ref{ng}),(\ref{neg}) and (\ref{nPG}) more explicitely in the PG phase at zero temperature:
\begin{eqnarray}\label{npg}
n
&=&
\int_{-4t}^{\mu_S} d\epsilon D(\epsilon) \, , \quad \quad \quad n_\epsilon =0 \, ,
\end{eqnarray}
\begin{eqnarray}\label{nepg}
n_{\overline{\epsilon}}
&=&\frac{1}{2}\sum_\pm
\int_{-4t}^{\mu_S} d\epsilon D(\epsilon)\frac{\epsilon}{4t} =\frac{2}{\pi^2} -
\frac{1}{\pi^2} \left( \frac{\mu_S}{4t}\right)^2\ln|\sqrt{e}\mu_S/16t|
\, .
\end{eqnarray}
Expanding Eq.(\ref{CN}) at low temperature, we find the CN free energy for the PG phase:
\begin{eqnarray}\label{fPG}
f^{PG}_{CN}
=\frac{U_{dR}}{2}n^2-4t_0n+4U_{d1}(2n^2- |n_{\overline{\epsilon}}|^2)-8U_{d2}n(n^2   -|n_{\overline{\epsilon}}|^2)+2U_{d3}(n^2 - |n_{\overline{\epsilon}}|^2 )^2 -\frac{\pi^2}{6} \frac{4|t|}{|\Delta_{PG}|}\frac{D(4t \mu/|\Delta_{PG}|)}{\beta^2} \,,
\nonumber \\
\end{eqnarray}
to be compared with the HF expression:
\begin{eqnarray}\label{fHF}
f^{HF}_{CN}
=\frac{U_{dR}}{2}n^2-4t_0 (n+n_\epsilon)+4U_{t1}n_\epsilon n+4U_{d1}(2n^2-n_\epsilon^2)
-(8U_{d2}n+4U_{t2} n_{\epsilon})(n^2   -n_\epsilon^2)
+2U_{d3}(n^2 - n_\epsilon^2)^2
 -\frac{\pi^2}{6} \frac{D(\mu)}{\beta^2} \,.
 \nonumber \\
\end{eqnarray}
In Eq.(\ref{fPG}), $n$ and $n_\epsilon$ are given by Eq.(\ref{npg}) and Eq.(\ref{nepg}) while in Eq.(\ref{fHF}), $n$ and $n_{\overline{\epsilon}}$ are given by Eq.(\ref{n}) and Eq.(\ref{ne}). Only the last terms depend quadratically on the temperature. The free energy reference with no hopping is:
\begin{eqnarray}\label{f0}
f^{0}_{CN}
=\frac{U_{dR}}{2}n^2-4t_0 n+8 U_{d1}n^2
-8U_{d2}n^3
+2U_{d3}n^4 \, .
 \nonumber \\
\end{eqnarray}

\subsection{Calculation of the antiferromagnetic potential $U_A$}

We consider the state with a doublon site $\vc{l}$ and a vacuum site $\vc{l} \pm \vc{1}_i$ in an adjacent site. With respect to these two sites there are six adjacent neighbouring sites.
The energy for this state is the sum of the doublon energy $U_d$, the second order energy for one hole of the doublon to occupy virtually  a neighbouring $p$ orbital site next to the $d$ vacuum site $\frac{2}{\epsilon_p-U_d+U_{pd}}$ or next to single occupied site
$\frac{6}{\epsilon_p-U_d+2 U_{pd}}$, and the second order energy of the neighbouring site hole to occupy virtually a neighbouring $p$ orbital site next to the $d$ vacuum site $\frac{3}{\epsilon_p}$, or next to the doublon site
$\frac{3}{\epsilon_p+2 U_{pd}}$. The total energy sum is
\begin{eqnarray}\label{UA1}
U_{A1}= U_{d} -t_{pd}^2\left(\frac{2}{\epsilon_p-U_d+U_{pd}} + \frac{3}{\epsilon_p+2 U_{pd}} + \frac{3}{\epsilon_p}+ \frac{6}{\epsilon_p-U_d+2 U_{pd}} \right) \,.
\end{eqnarray}
Next, we consider the situation with single occupancy everywhere. There are two energy  contributions of the $p$ site between  $\vc{l}$ and  $\vc{l} \pm \vc{1}_i$ and twelve for the other $p$ adjacent sites. The total energy is
\begin{eqnarray}\label{UA2}
U_{A2}=  -\frac{14 t_{pd}^2}{\epsilon_p + U_{pd}} \,.
\end{eqnarray}
At half-filling, we obtain the energy difference:
\begin{eqnarray}\label{UA}
U_{A}= U_{d} -t_{pd}^2\left(\frac{2}{\epsilon_p-U_d+U_{pd}} + \frac{3}{\epsilon_p+2 U_{pd}} + \frac{3}{\epsilon_p}+ \frac{6}{\epsilon_p-U_d+2 U_{pd}} -\frac{14}{\epsilon_p+U_{pd}}\right) \,.
\end{eqnarray}
For $U_{pd}=0$, we recover $U_A = U_{dR}$.

\end{widetext}

\bibliographystyle{apsrev4-1}
\bibliography{paperrefbcs}

\end{document}